\documentclass[rmp,twocolumn]{revtex4-1}

\usepackage{amsmath}
\usepackage{upgreek}
\usepackage{graphicx}

\graphicspath{{./python/}}

\begin{document}

\title{Laser-driven hole boring and gamma-ray emission in high-density plasmas}
\date{\today}
\author{E.~N.~Nerush}
\email{nerush@appl.sci-nnov.ru}
\author{I.~Yu.~Kostyukov}
\affiliation{Institute of Applied Physics of the Russian Academy of
Sciences, 46 Ulyanov St., Nizhny Novgorod 603950, Russia}
\affiliation{University of Nizhny Novgorod, 23 Gagarin Avenue, Nizhny
Novgorod 603950, Russia}

\begin{abstract}

Ion acceleration in laser-produced dense plasmas is a key topic of
many recent investigations thanks to its potential applications.
Indeed, at
forthcoming laser intensities ($I \gtrsim 10^{23} \text{
W}\,\text{cm}^{-2}$) interaction of laser pulses with plasmas can be
accompanied by copious gamma-ray emission. Here we demonstrate the
mutual influence of gamma-ray emission and ion acceleration during
relativistic hole boring in high-density plasmas with ultra-intense laser
pulses. If the gamma-ray emission is abundant, laser pulse reflection
and hole-boring velocity
are lower and gamma-ray radiation pattern is narrower than in the
case of low emission. Conservation of energy and momentum allows one
to elucidate the effects of the gamma-ray emission which are more
pronounced at higher hole-boring velocities.

\end{abstract}

\maketitle

\section{\label{sec:intro}Introduction}

Starting from the pioneering papers of Wilks and
Denavit~\cite{Wilks92, Denavit92}, showing that relativistically
intense laser pulses can bore holes in overdense plasmas and
accelerate ions by the radiation pressure, the radiation pressure
acceleration has been intensively studied and discussed.  E.g., the radiation
pressure acceleration was investigated in recent
experiments~\cite{Palmer11,Kar13}, relativistic hole boring was
studied theoretically~\cite{Schlegel09, Robinson09} and proposed for
fast ignition schemes~\cite{Naumova09}, it is shown by means of
numerical simulations that the radiation pressure acceleration is feasible
for acceleration of ions up to GeV energies~\cite{Esirkepov04, Ji14d},
see also Ref.~\cite{Macchi13}.

A number of planned laser facilities~\cite{Mourou14, Bashinov14}
aim to reach high field intensity in order to demonstrate a plenty of novel
phenomena such as electromagnetic
cascades~\cite{Bell08,Fedotov10,Nerush11c},
vacuum birefringence~\cite{King14}, and electron-positron pair
creation~\cite{Narozhny14}. Besides these effects requiring special
set-ups~\cite{Gonoskov13_2, Bashmakov14} a number of rough phenomena
should occur. E.g., incoherent synchrotron emission of hard photons will be an
inherent feature of extremely relativistic plasma
physics~\cite{Nakamura12, Ridgers12, Nerush14, Ji14a}. The latter phenomenon should
surely affect particle dynamics in the laser-driven hole boring.

Here we study the hole boring in high-density plasmas accompanied by
gamma-ray emission. A relativistic laser pulse (intensity $I \gtrsim 10^{18}
\text{ W}\, \text{cm}^{-2}$ for optical wavelengths) can drive plasmas
by its front~\cite{Wilks92}. The laser field pushes plasma electrons
which sweep ions by a longitudinal electric field in a thin shock-like layer of
charge separation~\cite{Macchi05,Schlegel09}. Some properties of the process
(front velocity, ion energy, etc.) can be found using momentum and
energy conservation laws in the nonrelativistic~\cite{Wilks92, Kruer75} and in the
relativistic cases~\cite{Robinson09, Naumova09}, and measured in
the experiments~\cite{Kar13}. If the laser intensity is
$I \gtrsim 10^{23} \text{ W}\, \text{cm}^{-2}$, plasma electrons can
efficiently emit high-energy photons.

One- and two-dimensional
simulations with radiation reaction force taken into account show that
the radiation losses decrease the hole-boring velocity, suppress
the filamentation process and affect
the electron and ion distributions~\cite{Schlegel09, Naumova09,
Tamburini10, Capdessus12}. It is generally believed that radiation losses do not
affect radiation pressure acceleration much if circularly polarized
pulses are used (particularly for thin targets)~\cite{Tamburini10, Tamburini12,
Chen11}. However, here we show that if laser intensity is extremely high, $I\sim 10^{25}~\text{W} \,
\text{cm}^{-2}$, high-energy photons can take away a great portion of laser
energy at least for thick targets (generation of hard photons in this
case can be stimulated by pair production, see
Sec.~\ref{sec:discussion}).

The approach that uses the radiation reaction force doesn't take into account
quantum nature of photon emission
that is important in some cases~\cite{Nerush11b, Duclous11}.
Moreover, radiation losses in quantum regime (when quantum
recoil is significant) cannot be
described by the concept of a classical force in
principle~\cite{Elkina11}.

Here we use the three-dimensional PIC code with emission of hard photons
taken into account by means of Monte Carlo method~\cite{Nerush14}.
Electron-positron pair production (photon decay) is also taken into account that allows us simulate the
hole boring process at intensities about $10^{25} \text{ W} \,
\text{cm}^{-2}$ and higher. Similar approach has been used in a number
of articles~\cite{Ridgers12, Ridgers13, Brady13b} where absence of
reflection, the modification of the ion spectrum and the influence of
gamma-ray emission on the energy budget are mentioned.  However, the
influence of incoherent emission on the hole boring was not the main
topic of these articles and was not covered in details.

Here we present detailed results of numerical simulations and
theoretical analysis that takes into account radiation losses during
the laser-driven hole boring and elucidates the key effects of gamma-ray
emission.

\section{Conservation of energy and momentum in laser-plasma
interactions}

\begin{figure}
\includegraphics{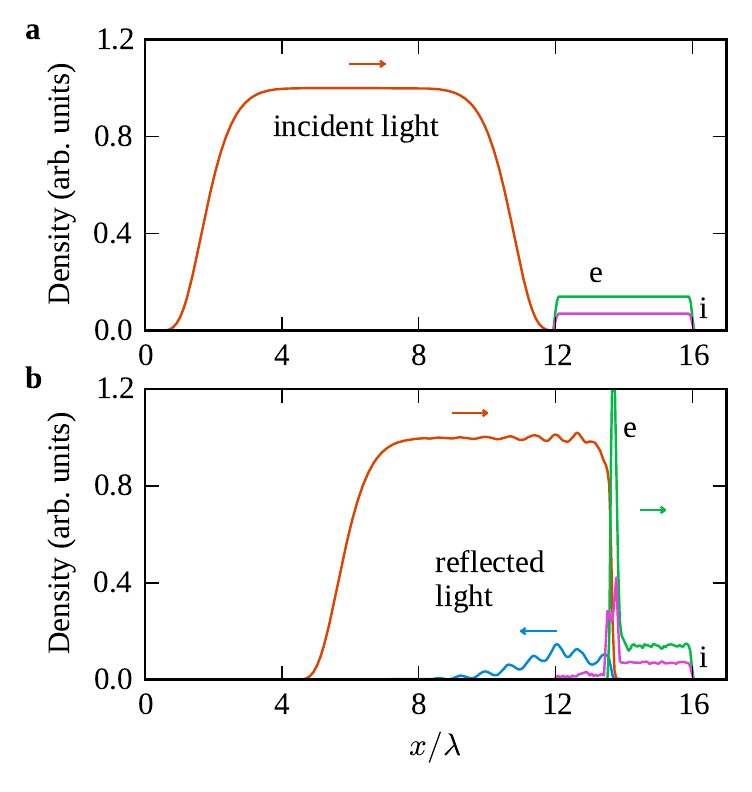}
\caption{\label{fig:shape}On-axis shape of the incident (orange) and reflected
(blue) laser pulses, electron density (green) and ion density
(violet) in a particle-in-cell simulation at time instances $t = 0$
(a) and $t = 4 \lambda/c$ (b).}
\end{figure}

To investigate the influence of gamma-ray generation on ion
acceleration we consider the interaction geometry where a laser pulse
is incident on a plasma slab producing reflected light, ion flow and
high-energy photons (Fig.~\ref{fig:shape}). According to
Ref.~\cite{Ji14a},  the ratio of the electron energy to the overall
energy in laser-plasma interactions decreases as the intensity increases, and for $I \gtrsim
10^{23} \text{ W}\, \text{cm}^{-2}$ the electron energy can become
negligible in the energy balance. Therefore, conservation of energy
and momentum is given as:
\begin{align}
\label{eq:energy}
( 1 - R ) S_l & = S_i + S_\gamma, \\
\label{eq:momentum}
( 1 + R ) S_l & = \Pi_i + S_\gamma \cos \varphi_0,
\end{align}
where $S_l$, $S_i$ and $S_\gamma$ are the laser pulse energy, the overall
ion energy and the overall energy of gamma-rays, respectively, $\Pi_i/c$
is the overall ion momentum, $\cos \varphi_0$ determines the ratio of
the longitudinal momentum of gamma-rays to their energy (i.e., $\varphi_0$
is the halfangle of the gamma-ray divergence), $R$ is the reflection
coefficient and $c$ is the speed of light. It follows from
Eqs.~(\ref{eq:energy})-(\ref{eq:momentum}) that
\begin{align}
\label{eq:pi-s}
\Pi_i - S_i & = 2 R S_l + ( 1 - \cos \varphi_0 ) S_\gamma, \\
\label{eq:pi+s}
\Pi_i + S_i & = 2 S_l - ( 1 + \cos \varphi_0 ) S_\gamma.
\end{align}

If $\Pi_i=0$, Eq.~(\ref{eq:pi-s}) is fulfilled only in the unrealistic case of
$R=0$, $S_i=0$ and $\varphi_0=0$. This means that $\Pi_i$ should be
greater than zero, hence the gamma-ray generation
is always accompanied by ion acceleration.

In addition, when the ion acceleration is accompanied by gamma-ray emission,
the former should be affected by the latter according to the momentum
end energy conservation. The effect of gamma-ray emission is quite
simple if the ions are ultrarelativistic.

In laser-foil interactions ions gain velocity almost equal to the
speed of light, for instance, in the laser-piston
regime~\cite{Esirkepov04}. In this regime light pressure pushes
a foil behaving as a mirror. The reflection coefficient tends to
zero due to relativistic motion of the foil~\cite{Esirkepov04}, and
gamma-rays are emitted primarily in the forward direction ($\varphi_0
\simeq 0$) because of light aberration and relativistic
Doppler effect~\cite{Nerush14, Ridgers12}. These features of
laser-plasma interaction apparently are also inherent for ultrarelativistic
hole-boring. Furthermore, for ultrarelativistic ions $\Pi_i
\simeq S_i$, hence Eq.~(\ref{eq:pi-s}) is naturally fulfilled, and
Eq.~(\ref{eq:pi+s}) yields $S_i \simeq S_l - S_\gamma$. Therefore,
the gamma-ray emission just lowers the ion acceleration rate if ions are
ultrarelativistic.

\section{\label{sub:mr}Energy and momentum balance in laser-driven hole boring}

The laser-driven hole boring is a well-known interaction regime that can
be roughly described as ion acceleration by a kick from a shock moving together with the laser
pulse front with the speed $v_{hb}$~\cite{Wilks92, Schlegel09,
Robinson09}.
Rebounded ions overtake the front and, since all of them move with the same
velocity $v_r$, they form a quasimonoenergetic distribution. In fact,
ions in this regime are accelerated by the electric field induced
by the laser pulse through electron-ion separation~\cite{Schlegel09,Nerush14}.
Every
rebounded ion passes the same track in the field, hence, all of them
gain the same energy. This idealized picture allows one to
find the speed of the laser pulse front, $v_{hb}$, as demonstrated in
Sec.~\ref{sec:lgre} for almost absent gamma-ray emission and in
Sec.~\ref{sec:agre} for abundant gamma-ray emission.

\subsection{\label{sec:lgre}Low gamma-ray emission}

The conservation laws Eqs.~(\ref{eq:energy}), (\ref{eq:momentum}) can be
written not only for the overall energy and momentum, but also for
the energy and momentum flow rates per unit area of the shock:
\begin{align}
\label{eq:energy-low}
\frac{ ( 1 - R ) ( c - v_{hb} ) }{ 4 \pi } \left| \mathbf E \times
\mathbf B \right | & = n_i v_{hb} M c^2 ( \gamma_r - 1 ), \\
\label{eq:momentum-low}
\frac{ ( 1 + R ) ( c - v_{hb} ) }{ 4 \pi } \left| \mathbf E \times
\mathbf B \right | & = n_i v_{hb} M c v_r \gamma_r,
\end{align}
where $\mathbf E$ and $\mathbf B$ are electric and magnetic fields
of the laser pulse, $n_i$ is the initial ion density, $M$ is the
ion mass and $\gamma_r = c\left( c^2 - v_r^2 \right)^{-1/2}$ is the
Lorentz factor of the rebounded ions. It follows directly from these
equations that
\begin{align}
\label{eq:R}
R & = \sqrt{ \frac{ c - v_r }{ c + v_r } },\\
\label{eq:gammar}
\frac{ c - v_{hb} }{ 2 \pi } \left| \mathbf E \times \mathbf B \right|
& = M c n_i v_{hb} \left[ \gamma_r ( c + v_r ) - c \right].
\end{align}

The velocity and the Lorentz factor of the accelerated ions can be
found from the assumption of perfectly elastic rebound in the
reference frame co-moving with the front of the laser pulse:
\begin{equation}
\label{eq:vr}
v_r = \frac{ 2 c^2 v_{hb} }{ c^2 + v_{hb}^2 }, \qquad \gamma_r = \frac{ c }{
\sqrt{ c^2 - v_r^2 } } = \frac{ c^2 + v_{hb}^2 }{ c^2 - v_{hb}^2 }.
\end{equation}
From these equations together with Eqs.~(\ref{eq:R}) and (\ref{eq:gammar}) we get, first, $R = ( c - v_{hb} )/( c + v_{hb}
)$. This formula corresponds to a complete laser light reflection in the
reference frame
co-moving with the front of the laser pulse. Second,
\begin{equation}
\label{eq:vhb}
v_{hb} = \frac{ c }{ 1 + \mu }, \qquad \mu = \sqrt{ \frac{ 4 \pi c^2
M n_i }{ \left| \mathbf E \times \mathbf B \right| } } = \frac{ 1 }{ a_0 } \sqrt{ \frac{ M n_i }{ m n_{cr} } },
\end{equation}
that matches with the results obtained in Refs.~\cite{Schlegel09,
Robinson09} by a slightly
different approach. Here $n_{cr} = m \omega^2 / 4 \pi e^2$ is the
critical density, $\omega$ is the laser cyclic frequency, $m$ and $e>0$
are the electron mass and the electron charge magnitude, respectively,
$a_0 = e E_0 / m c \omega$ is the normalized amplitude of the incident
laser pulse, $E_0$.

\begin{figure}
\includegraphics{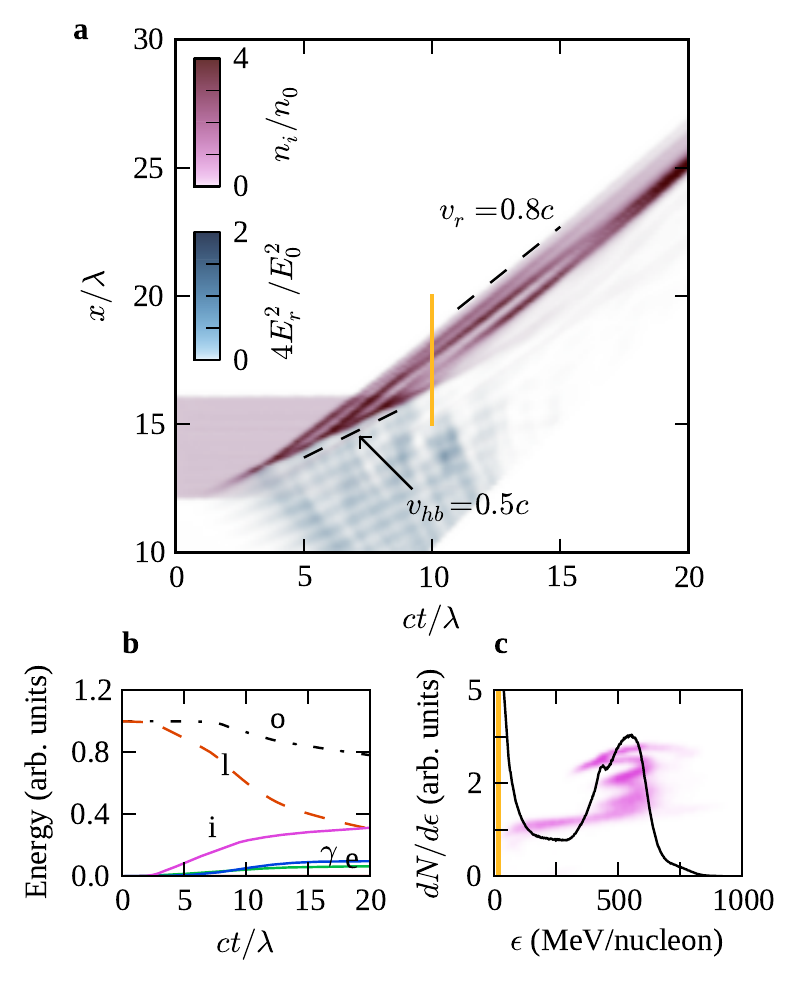}
\caption{\label{fig:nxt-low}The results of a numerical simulation of a
circularly polarized laser pulse ($a_0=500$) in normal incidence on a
$4 \text{ }\upmu \text{m}$-thick He slab (initial ion density is $n_0
= 35 n_{cr}$). (a) On-axis ion density normalized to the initial ion
density and intensity of the reflected light; $E_r^2 = \left(
\mathbf E - \mathbf B \times \mathbf e_0 \right)^2/4$ is the squared
electric field of the reflected wave, $E_0$ is the amplitude of the
incident wave and $\mathbf e_0$ is the unit vector of the propagation
direction of the incident pulse. (b) The overall, laser, ion and
electron energy, and the energy of gamma-rays during the interaction.
The decline in the overall energy corresponds to the exit of the
reflected light from the simulation box. (c) Spectrum of all ions and
the distribution of on-axis ions in the $\epsilon-x$
phasespace at $t = 10 \lambda/c$. Here $\epsilon = Mc^2 (\gamma - 1)$
is the kinetic energy of an ion.}
\end{figure}

Results of a three-dimensional particle-in-cell simulation with
incoherent emission taken into account illustrate
well hole boring with an intense laser pulse in a dense plasma
(Figs.~\ref{fig:shape}, \ref{fig:nxt-low}). The simulation box is $
17.5 \lambda \times 25 \lambda \times 25 \lambda $ corresponding to
the grid size $ 670 \times 125 \times 125 $ with the cell size $ 0.026
\lambda \times 0.2 \lambda \times 0.2 \lambda $. The time step is $
0.019 \lambda/c $, where $\lambda$ is the laser wavelength. The
initial number of
quasiparticles is $3.8 \times 10^7$ corresponding to $ 16 $ particles per cell in
the unperturbed plasma.
In the simulation a quasi-rectangular ($12 \lambda \times 24 \lambda \times 24 \lambda$)
circularly polarized laser pulse of intensity $7 \times 10^{23} \text{
W} \, \text{cm}^{-2}$ ($\lambda = 1 \text{ } \upmu \text{m}$, $a_0 = 500$)
interacts with a He slab ($n_e = 2 n_i = 7.8 \times 10^{22} \text{
cm}^{-3}$, $n_i/n_{cr} = 35$). Helium
becomes fully ionized, then electrons are pushed by $\mathbf v
\times \mathbf B$ force resulting charge separation that creates accelerating field.
For the given parameters $\mu = 1$ and $v_{hb} = 0.5 c$ that is in fairly good
agreement with the numerical results [Fig.~\ref{fig:nxt-low}(a)].
The overall ion energy grows linearly with time while the laser pulse
front
is passing the slab
[$3 \lambda/c \lesssim t \lesssim 9 \lambda/c$, see Fig.~\ref{fig:nxt-low}(b)].
However, the rebound of ions is not regular in time and the resulting ion spectrum is not strictly
monoenergetic [Fig.~\ref{fig:nxt-low}(c); on the broadening of the ion
spectrum and modulations in the phasespace see also
Refs.~\cite{Robinson09, Schlegel09, Macchi13}].

\subsection{\label{sec:agre}Abundant gamma-ray emission}

According to Eq.~(\ref{eq:vhb}) that doesn't take into account
high-energy photons, the hole-boring velocity remains the same if $M
n_i$ (i.e. mass density) is increased proportionally to the laser intensity. If the hole-boring velocity is
non-relativistic, the gamma-ray
generation efficiency is low and doesn't grow with simultaneous
increase of the intensity and the density~\cite{Nerush14}. However, the energy taken away by
gamma-rays increases sharply with the intensity and can be of the order of the
initial laser pulse energy in the case of relativistic hole boring.

\begin{figure}
\includegraphics{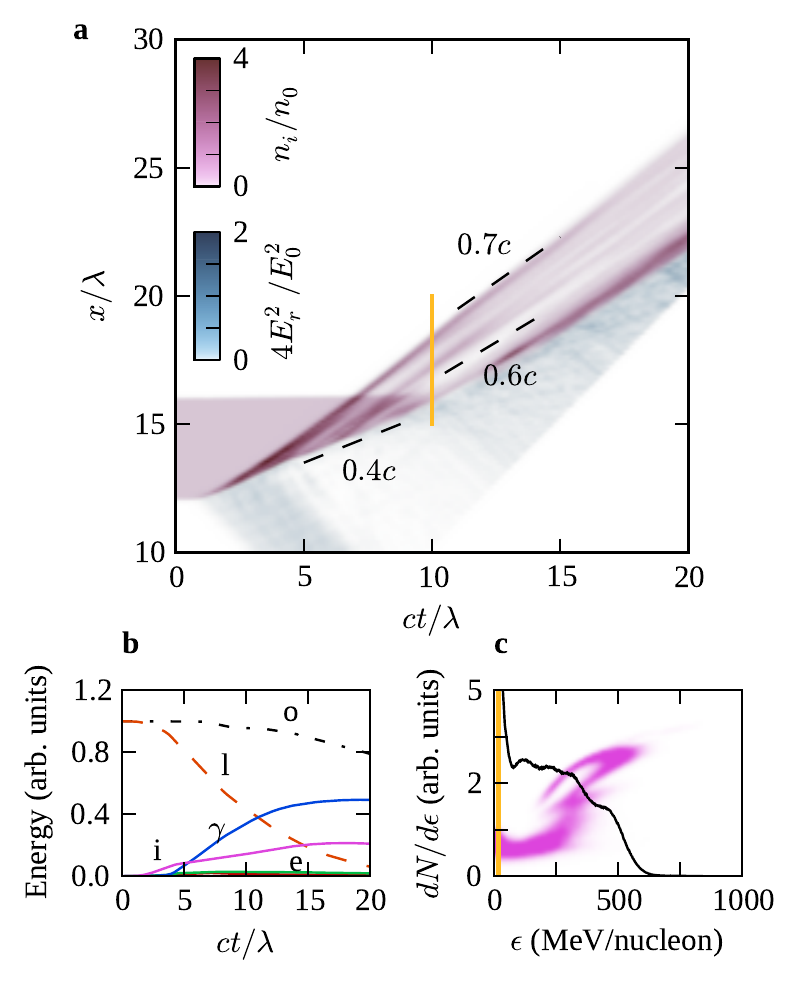}
\caption{\label{fig:nxt-med} The results of a numerical simulation of a
circularly polarized laser pulse ($a_0=1840$) in normal incidence on a
$4 \text{ }\upmu \text{m}$-thick diamond foil (initial ion density is
$n_0 = 158 n_{cr}$). (a) On-axis ion density and intensity of the
reflected light. The sharp front of the reflected field distribution
at $t \gtrsim 10$ is caused by a boundary of the simulation box. (b)
The overall, laser, ion and electron energy, and the energy of
gamma-rays during the interaction.  (c) Spectrum of all ions and the
distribution of on-axis ions in $\epsilon-x$
phasespace at $t = 10 \lambda/c$.}
\end{figure}

\begin{figure}
\includegraphics{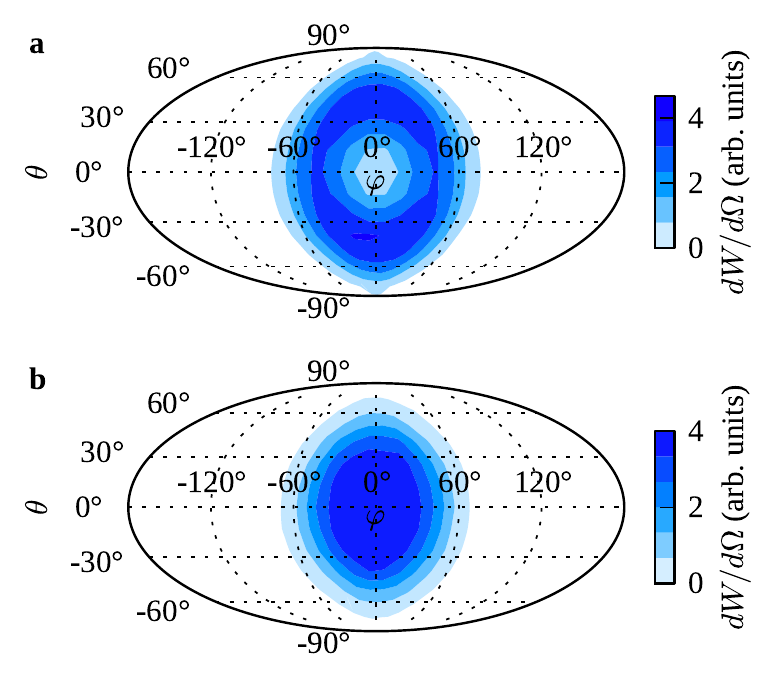}
\caption{\label{fig:rp}The Mollweide projection of the gamma-ray
radiation pattern, i.e. the angular distribution of the emitted
energy, in normal incidence of a laser pulse ($a_0=1840$) on a $4
\text{ } \upmu \text{m}$-thick diamond foil at $t=5 \lambda/c$ (a) and
$t = 10 \lambda/c$ (b). The longitude $\varphi$ and
the latitude $\theta$ are introduced so that the point $\varphi=0$,
$\theta=0$ corresponds to the propagation direction of the incident
laser pulse.}
\end{figure}

PIC simulation (Fig.~\ref{fig:nxt-med}; the simulation box is $
17.5 \lambda \times 25 \lambda \times 25 \lambda $, the cell size is $ 0.007
\lambda \times 0.17 \lambda \times 0.17 \lambda $, the time step is $ 0.005 \lambda/c $, the initial number of
quasiparticles is $2 \times 10^8$ that is barely enough to resolve high-energy tails in the electron distribution, thus massive simulations are desirable in the future) shows that in the interaction
of a $9.3 \times 10^{24} \text{ W} \, \text{cm}^{-2}$ ($a_0 = 1840$) laser pulse with
a diamond foil ($n_e = 6 n_i = 1.1 \times 10^{24} \text{ cm}^{-3}$,
$n_i/n_{cr} = 158$)
the fraction
of energy carried away by gamma-rays is $2.3$ times larger than that
of ions [Fig.~\ref{fig:nxt-med}(b)]; in the previous case, for a He slab and a laser pulse of $a_0=500$, this ratio is only
$0.3$. Though $\mu$ is the same in both cases, such sizeable
energy losses, caused by incoherent synchrotron emission in the second case, considerably affect the
interaction process. Primarily, the reflection coefficient, the
hole-boring velocity and the ion spectrum are involved.

To determine the effect of radiation losses we introduce them
into Eqs.~(\ref{eq:energy-low}) and (\ref{eq:momentum-low}):
\begin{multline}
\label{eq:energy-ab}
\frac{ ( 1 - R ) \left( c - v_{hb} \right) }{ 4 \pi } \left| \mathbf E \times
\mathbf B \right| =\\ n_i v_{hb} M c^2 \left( \gamma_r - 1 \right) +
S_\gamma,
\end{multline}
\begin{multline}
\label{eq:momentum-ab}
\frac{ ( 1 + R ) \left( c - v_{hb} \right) }{ 4 \pi } \left| \mathbf E
\times \mathbf B \right| =\\ n_i v_{hb} M c
v_r \gamma_r + S_\gamma \cos \varphi_0.
\end{multline}
From these equations the reflection coefficient can be found as
the function of the rebound speed $v_r$, the emission angle $\varphi_0$ and
\begin{equation}
\eta_{\gamma i}= S_\gamma/n_i v_{hb} M c^2 ( \gamma_r - 1 ),
\end{equation}
the fraction of
the energy carried away by gamma-rays to the energy carried away by ions, as follows:
\begin{align}
\label{eq:R-med}
R & = \sqrt{ \frac{ c - v_r }{ c + v_r } } \; \frac{ 1 - ( 1 -
\cos \varphi_0 ) \eta_{\gamma i} f^- }{ 1 + ( 1 + \cos \varphi_0 ) \eta_{\gamma i} f^+ }, \\
f^\pm & = \frac{ 1 }{ \sqrt{ c \pm v_r } } \; \frac{ c - \sqrt{ c^2 - v_r^2
}}{ \sqrt{ c + v_r } - \sqrt{ c - v_r } }.
\end{align}
If gamma-ray emission is abundant, $\eta_{\gamma i} \gg 1$ (for
instance, from the slope of the curves in
Fig.~\ref{fig:nxt-med}(b) we obtain $\eta_{\gamma i} \simeq 5$).
The function $f^+$ increases from $0$ to $1/2$ and the function $f^-$ changes from $0$ to $\infty$ if $v_r$ grows from
$0$ to $1$. Thus, if the hole-boring is relativistic, the products
$\eta_{\gamma i} f^+$ and $\eta_{\gamma i} f^-$ become greater than unity and the
reflection coefficient in this case tends to
zero. Furthermore, if $\eta_{\gamma i} f^- \gg 1$, the gamma-rays are emitted mostly in the
forward direction ($\cos \varphi_0 \simeq 1$), otherwise
Eq.~(\ref{eq:R-med}) yields $R<0$. As $f^+$ and $f^-$ increase with the
increase of $v_r$, the decrease of the reflection coefficient and
the narrowing of the gamma-ray radiation pattern caused by the radiation
losses become more pronounced at higher hole-boring velocities.

The effects following from Eq.~(\ref{eq:R-med}) are in fairly good
agreement with
the simulation results presented in Fig.~\ref{fig:nxt-med}.  Most portion of gamma-rays are emitted at
$5 \lambda/c \lesssim t \lesssim 10 \lambda/c$, and the reflection
coefficient is very low
during this time interval [Fig.~\ref{fig:nxt-med}(a)]. In addition,
the gamma-ray radiation pattern at $t = 5 \lambda/c$ is conical with
$\varphi_0 \simeq 45^\circ$, and the radiation pattern at $t = 10
\lambda/c$ is rather spotlight-like (Fig.~\ref{fig:rp}). The
difference between these patterns can be interpreted as follows.  For
$t \lesssim 5 \lambda/c$ the gamma-ray generation efficiency is low,
hence radiation losses do not affect the laser-plasma interaction much
and conical radiation pattern is formed [the radiation pattern formed
in the incidence of a laser pulse with $a_0=500$ on a He slab (see
Sec.~\ref{sec:lgre}) is almost the same]. Most part of gamma-rays
is emitted during the time interval $5 \lambda/c \lesssim t \lesssim
10 \lambda/c$, that affects the laser-plasma interaction and leads to the
narrowing of the gamma-ray radiation pattern [Fig.~\ref{fig:rp}(b)].

We assume that when a laser pulse bores a hole in a plasma, a potential barrier is
created at the front of the laser pulse in the case of abundant
gamma-ray emission also. Rebound of ions from the
moving potential barrier is elastic and, in the reference frame
co-moving with the barrier, value of the ion velocity doesn't change
after the rebound. Thus, in the laboratory reference frame the velocity of
the laser pulse front and the velocity of the rebounded ions are related
according to Eq.~(\ref{eq:vr}). Together with
Eqs.~(\ref{eq:energy-ab})-(\ref{eq:momentum-ab}) this yields:
\begin{equation}
\frac{ ( c - v_{hb} ) }{ 4 \pi } \left| \mathbf E \times \mathbf B
\right| = \frac{ Mc^2 n_i v_{hb}^2 }{ c - v_{hb} } + \frac{S_\gamma ( 1 + \cos \varphi_0 )}{ 2 },
\end{equation}
and the velocity of the laser pulse front, $v_{hb}$, is determined by
Eq.~(\ref{eq:vhb}) with $\tilde \mu$ substituted for $\mu$:
\begin{align}
\label{eq:tildemu}
v_{hb} = \frac{ c }{ 1 + \tilde \mu }, & \qquad
\tilde \mu = \frac{ \mu }{ \sqrt{ 1 - \sigma_{\gamma l} }}, \\
\mu = \frac{1}{a_0} \sqrt{ \frac{ M n_i }{ m n_{cr} } }, & \qquad \sigma_{\gamma l} = \frac{ 2 \pi ( 1 + \cos \varphi_0 ) S_\gamma }{ (
c - v_{hb} ) \left| \mathbf E \times \mathbf B \right| }.
\end{align}

At high radiation losses, when $R \simeq 0$ and $\cos \varphi_0 \simeq 1$,
$\sigma_{\gamma l}$ is approximately equal to the
gamma-ray generation efficiency, i.e., the ratio of the emitted
gamma-ray energy, $\tau S_\gamma$, to the absorbed laser pulse energy,
$\tau (c-v_{hb}) \left| \mathbf E \times \mathbf B \right| / 4 \pi$, where $\tau $ is the interaction time.  From
the slope of the curves in Fig.~\ref{fig:nxt-med}(b) for $5 \lambda/c
\lesssim t \lesssim 10 \lambda/c$ we obtain $\sigma_{\gamma l} \simeq
0.7$, hence Eqs.~(\ref{eq:tildemu}) and (\ref{eq:vr})
yield $\tilde \mu = 1.8$, $v_{hb} = 0.36 c$ and $v_r = 0.6 c$
($\epsilon = 270 \text{ MeV} / \text{nucleon}$). These analytical
results describes well ion dynamics in the numerical simulation
(Fig.~\ref{fig:nxt-med}). However, the ion distribution in $\epsilon - x$
phasespace consist not only of vertical part with $\epsilon \simeq 250
\text{ MeV} / \text{nucleon}$, but also of parts with higher and lower
energy. Fig.~\ref{fig:nxt-med}(a) demonstrates that ions with $v_r
\gtrsim 0.6 c$ are accelerated when the gamma-ray generation is not yet
efficient ($t \lesssim 5 \lambda/c$) and ions with $v_r \lesssim 0.6
c$ are accelerated when the laser front leaves the foil that is accompanied
by a decrease in the accelerating field ($t \simeq 10 \lambda/c$).
The amount of low-energy ions in the case of $a_0 = 500$ laser pulse
and He slab (Sec.~\ref{sec:lgre}) is much lower than in the
considered case. This indicates, possibly, much longer acceleration time in the
case of abundant emission. However, here we do not consider in detail how (from the microscopic point of view) efficient
gamma-ray generation is connected with the decrease in ion
acceleration rate and in reflection coefficient, and how it is
connected with
narrowing of the radiation pattern. We limit ourselves to
demonstration of these general effects of gamma-ray emission.

\section{\label{sec:discussion}Discussion}

\begin{figure}
\includegraphics{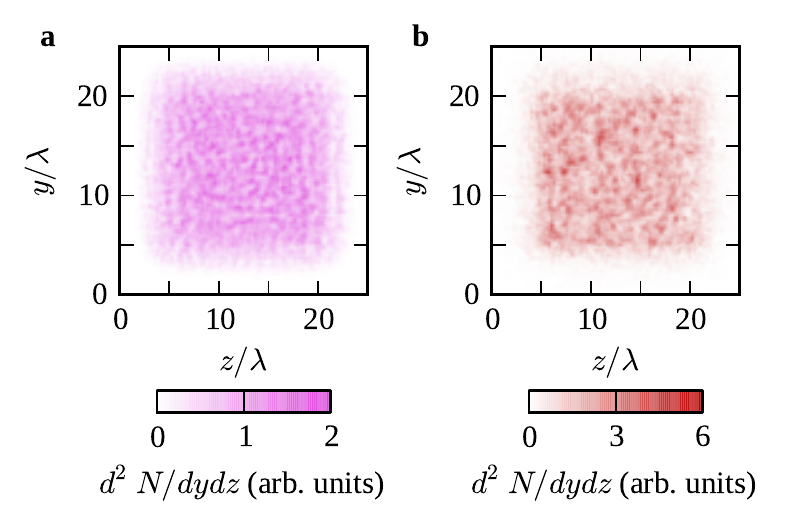}
\caption{\label{fig:ions}Distribution of ions with $\epsilon > 250
\text{ MeV}/\text{nucleon}$ (a) and positrons along the plane of the
shock generated in normal incidence of a circularly polarized laser
pulse ($a_0 = 1840$) on a diamond foil. The distributions are
normalized to the average value of the ion distribution.}
\end{figure}

If the incoherent photon emission is absent,
energy of a laser pulse that bores a hole in a dense plasma is
transmitted mostly to ions and reflected light. In this case proportional
increase of the laser intensity and the plasma mass density doesn't change the
interaction pattern (hole-boring velocity, velocity of ions, etc.).
However, at high intensities efficient generation of hard photons
occurs that changes energy and momentum balances. First, as a great part of
the laser energy is transmitted to hard photons, the energy of the
reflected light drops together with the hole-boring velocity and the energy of
rebounded ions. Second, the gamma-ray radiation
pattern should be quite narrow and directed
along the initial direction of the laser pulse, because a great part of the laser momentum
is carried away by gamma-rays and is not taken away by the ions. We show that the drop
in the reflection coefficient and the narrowing of the gamma-ray radiation
pattern caused by the losses are more pronounced for higher hole-boring velocities. In the
opposite, the kinetic energy of the rebounded ions $\gamma_r - 1 = 2
/ ( 2 \tilde \mu + {\tilde \mu}^2 )$ is more sensitive to the
radiation losses if $\tilde \mu \gg 1$, i.e. at low hole-boring velocities.
Nonetheless, at high hole-boring velocities radiation losses also
influence on the ion acceleration and can reduce the acceleration
rate.

Three-dimensional
particle-in-cell simulations with incoherent emission of hard photons
and pair production taken into account
agree well with the proposed picture that follows from the energy and
momentum conservation laws. However, numerical simulations reveal a number of phenomena not taken
into account by the analytical model. First, even in the case of low
gamma-ray emission the beam of accelerated ions is not strictly
homogeneous, and the ion spectrum is not strictly monoenergetic
(Fig.~\ref{fig:nxt-low}). Second, in the ultrahigh-intensity case
the ion acceleration rate changes with time leading to
a broad ion spectrum (Fig.~\ref{fig:nxt-med}). The broadening of the ion
spectrum in the low-emission case is caused by the piston
oscillations~\cite{Robinson09, Schlegel09, Macchi13} that lead to the
modulations in the phasespace [Fig.~\ref{fig:nxt-low}(c)] and
modulations in the reflected light [Fig.~\ref{fig:shape}(b)]. At the
same time, the broad ion spectrum and modulations in the phasespace
[Fig.~\ref{fig:nxt-med}(c)] in ultrahigh-intensity case are caused by
the radiation losses that are correlated with pair production (see below).

The beam of the rebounded ions is propagating through the unperturbed plasma
that causes Weibel-type instability and beam
filamentation~\cite{Fox13}, filamentation of the laser pulse
front~\cite{Naumova09, Palmer12} also occurs. Filamentation of the ion beam and the laser
pulse front is more pronounced in the case of
low gamma-ray emission.
In the given numerical simulations ion
filaments have no time to coalesce, however, for a thick foils
filamentation of the ion
flow and magnetic field generation can significantly change the
ion distribution function, as happens in plasmas of gamma-ray
burst outflows~\cite{Silva03, Bret14}

In the case of abundant gamma-ray
emission hard photons decay and
create electron-positron pairs in
strong laser and plasma fields. Numerical simulations
show that in the case of ultra-intense laser
pulse pair production is crucial for efficient gamma-ray
generation. For the simulation with $a_0 = 1840$
the number of produced positrons is about the initial number of
electrons in the simulation box, and the number of hard photons
produced by the positrons is about the number of hard photons produced by electrons. The possible explanation is that the
generated positrons can stay near the shock for a long time and
generate gamma-rays efficiently.
If pair production is not included in PIC simulation,
the gamma-ray generation efficiency, the hole boring velocity and the
ion spectrum are close to that in the case of much less intense laser
pulse ($a_0 = 500$) and He slab. Thus, the feedback between
processes of quantum electrodynamics and plasma physics~\cite{Nerush11a, Ridgers13,
Ji14b} is
evident in this case. The positron dynamics
during hole boring with ultra-intense laser pulses is studied in
Ref.~\cite{Kirk13}, however,
a vacuum gap assumed
there is absent in our simulations; moreover, filamentation of the positron beam
occurs (Fig.~\ref{fig:ions}). This indicates that further investigations are needed.
Nevertheless, pair plasma does not directly involved in energy and
momentum balances and just converts laser energy into gamma-ray
energy. Hence, the analysis presented here is also valid in the case of abundant
pair production.

In conclusion, the influence of radiation losses on a hole boring with
ultra-intense laser pulses is considered by means of
three-dimensional simulations and theoretical analysis. Conservation of
momentum and energy allows one to understand how the losses affect
hole boring properties for any mechanism of
the incoherent photon emission.

\begin{acknowledgments}
This work has been supported in part by the Government of the Russian
Federation (Project No. 14.B25.31.0008) and by the Russian Foundation
for Basic Research (Grants No 13-02-00886, 15-02-06079).
\end{acknowledgments}

\bibliography{hb-with-gre}
\end{document}